\documentclass[conference]{IEEEtran}
\IEEEoverridecommandlockouts
\usepackage{cite}
\usepackage{amsmath,amssymb,amsfonts}
\usepackage{algorithmic}
\usepackage{graphicx}
\usepackage{textcomp}
\usepackage{xcolor}
\usepackage{hyperref}
\usepackage{physics}
\def\BibTeX{{\rm B\kern-.05em{\sc i\kern-.025em b}\kern-.08em
    T\kern-.1667em\lower.7ex\hbox{E}\kern-.125emX}}

\begin{document}

\title{A Quantum Platform for Multiomics Data}

\author{
\IEEEauthorblockN{Michael Kubal}
\IEEEauthorblockA{\textit{Coherent Computing Inc} \\
\textit{Chicago, Illinois} \\
}

\and

\IEEEauthorblockN{Sonika Johri}
\IEEEauthorblockA{\textit{Coherent Computing Inc\thanks{Corresponding author email: sjohri@coherentcomputing.com}} \\
\textit{Cupertino, California}
}
}

\maketitle

\thispagestyle{plain}
\pagestyle{plain}

\begin{abstract}
The complexity of biological systems, governed by molecular interactions across hierarchical scales, presents a challenge for computational modeling. While advances in multiomic profiling have enabled precise measurements of biological components, classical computational approaches remain limited in capturing emergent dynamics critical for understanding disease mechanisms and therapeutic interventions. Quantum computing offers a new paradigm for addressing classically intractable problems, yet its integration into biological research remains nascent due to scalability barriers and accessibility gaps. Here, we introduce a hybrid quantum-classical machine learning platform designed to bridge this gap, with an encode-search-build approach which allows for efficiently extracting the most relevant information from biological data to \underline{encode} into a quantum state, provably efficient training algorithms to \underline{search} for optimal parameters, and a stacking strategy that allows one to systematically \underline{build} more complex models as more quantum resources become available. We propose to demonstrate the platform's utility through two initial use cases: quantum-enhanced classification of phenotypic states from molecular variables and prediction of temporal evolution in biological systems. 

\end{abstract}


\section{Introduction}
The human body has about 36 trillion cells, most containing DNA with 3 billion base pairs that encode the instructions for constructing tens of thousands of protein machines responsible for orchestrating the complex balance of energy, biochemical reactions and information that make life possible \cite{Hatton2023-ab}. With today's technology, we have the capability to measure the quantity, structure and characteristics of various molecular entities with staggering precision within various compartments and systems of the biological hierarchy. The cures for many diseases are now within reach as we understand many of the components, their interactions, and points of failure involved in diseases. To efficiently and ethically test our understanding and possible interventions we need scalable models of how these entities interact with each other to produce biological outcomes. 

Today, we have many useful computational models available for different systems from individual protein-ligand interactions to organism-scale metabolism \cite{Ponce-de-Leon2023}. However, relying solely on classical computing limits our ability to model nonlinear, long-range effects shaping emergent behaviors in complex, even chaotic systems.

Quantum computers can be applied to many classically challenging computational problems in biology such as classification (cell state, phenotype, disease state, etc), optimization (pathway regulation, balancing stability and growth) and chemistry modeling (molecular binding, enzyme kinetics). Here we focus on quantum learning models (QLM) which offer a new paradigm with the ability to model beyond-classical correlations in data and efficiently train models that have a large number of parameters with relatively modest energy usage. 

In this conceptual paper, we outline a hybrid quantum-classical platform that achieves the following: 1) encodes high-dimensional multiomic/multimodal data into a form that can be effectively loaded into quantum states 2) enhances existing machine learning frameworks used by computational biological models with quantum approaches to more comprehensively explore the model's parameter/solution space 3) uses the recently proposed quantum subnet-initialization technique \cite{johri2025} to organize the framework so that smaller quantum models can be built today that can be seamlessly aggregated into larger models as the number of qubits increases.

\section{Quantum Omics Platform}
\begin{figure*}
    \centering
    \includegraphics[width=\linewidth]{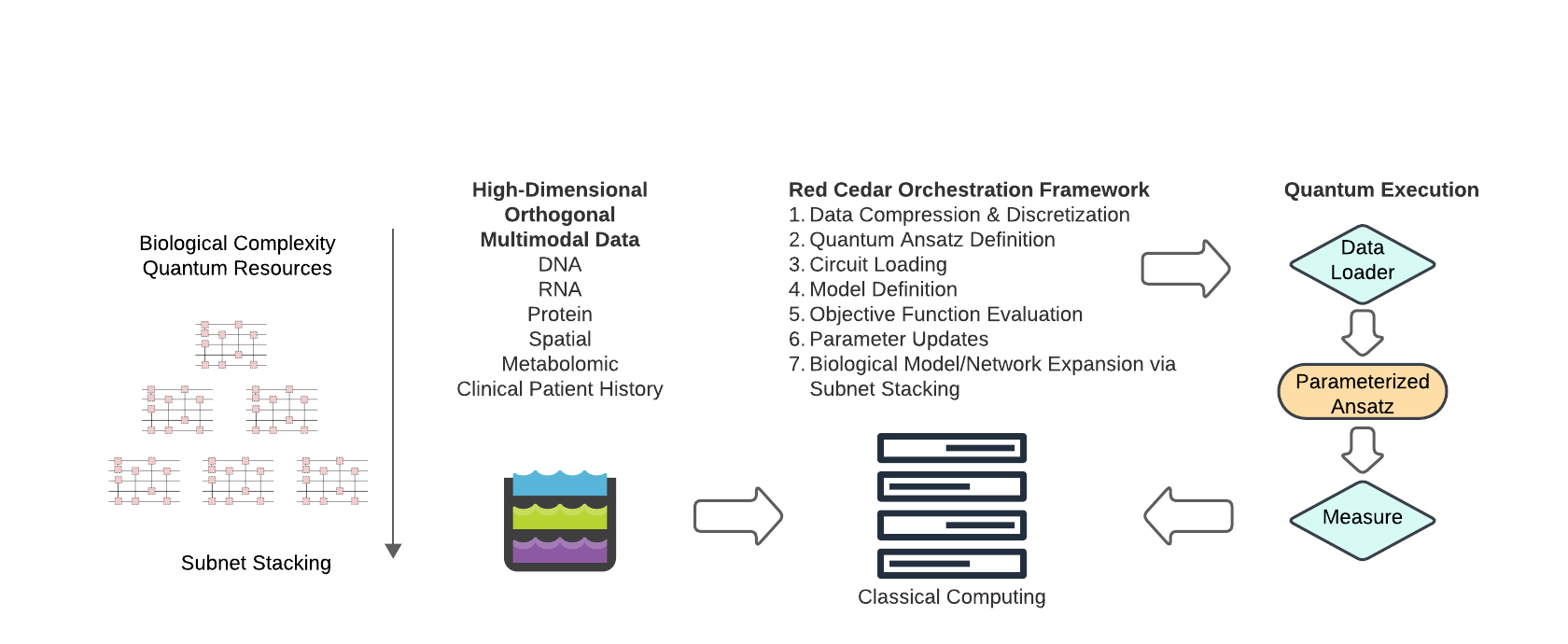}
    \caption{Schematic description of the computational workflow. The quantum execution can consist of several ansatzes that are then post-processed for inclusion in the model. The model can be a canonical machine learning model like logistic regression, a mechanistic ODE-driven pathway, or a high resolution molecular interaction model.}
    \label{fig:workflow}
\end{figure*}

The potential of quantum computing in analyzing omics data has previously been considered at an abstract level \cite{maniscalco2022quantumnetworkmedicine, basu2023quantumenabled}. Numerical studies on simulators have been confined to examples with very few qubits and/or low entanglement, such as quantum generative models with 2 and 4 qubits in \cite{li2021_drug_discovery}, and a single layer of entangling gates in \cite{Li2-21_quantum_gen}. This reflects the challenges in scaling up the training of machine learning models to the point where they become useful in the real world.

Here we propose to leverage recent advancements in techniques for scalable machine learning \cite{johri2025} to build an end-to-end hybrid quantum-classical platform that can process real omics data and is integrated with state-of-the-art quantum simulators and hardware. Our platform will have an extensible architecture that can be customized to different applications and can support increasingly more sophisticated models as quantum resources, such as the number of qubits and their gate fidelities, improve.

This platform will utilize Coherent Computing's Red Cedar software framework (Fig. \ref{fig:workflow}) \cite{johri2025}, which supports the development and training of quantum machine learning models. The framework follows an encode-search-build approach which allows for efficiently extracting the most relevant information from data to \underline{encode} into a quantum state, provably efficient training algorithms to \underline{search} for optimal parameters, and a stacking strategy that allows one to systematically \underline{build} more complex models as more quantum resources become available. In addition to the algorithms, the framework also features custom optimization and transpilation to a number of backends that enable optimal use of available quantum resources. 

Our quantum omics platform will seamlessly integrate biological data sources, streamline data preparation and loading, standardize the construction of complex quantum circuits, enable dynamic workflows, and generate performance benchmarks. These capabilities are critical for applying quantum machine learning models to biological challenges. By providing intuitive utilities and a structured development environment, this will drive broader adoption and accelerate innovation. Our platform will bridge the gap between the vast repositories of multi-omics data, the expanding quantum hardware ecosystem, and bioinformaticians and data scientists well-versed in machine learning but hindered by the absence of an accessible quantum framework. Unlike existing quantum software frameworks such as Qiskit \cite{qiskit2024} which require deep knowledge of quantum computing concepts, we will provide an end-to-end workflow—from raw molecular data to quantum-enhanced analysis—delivering actionable biological insights.

There are many potential biological use cases in genomics and beyond to which the encode-search-build approach can be applied. The first end-to-end use case to be enabled in our platform is predicting a phenotype (or biological outcome/state) given a set of molecular variables (such as genetic variants/biomarkers, protein/RNA expression levels, methylation status derived from sequencing readouts and spatial imaging). This effort will replace a classical machine learning approach such as logistic regression with a quantum classification algorithm. The second involves predicting how these variables evolve over time in a well-defined biological context such as an enzymatic and/or signaling pathway, cell or ultimately an entire organ or organism. This will replace a classical differential equation solver which gets inaccurate over long timescales when there are non-linear relationships between variables.

\section{Algorithmic Approach}

Quantum algorithms provide an alternate paradigm to model correlations in many-body systems that may not be efficient to simulate classically. We consider a biological system, such as a cell, characterized by multiple correlated variables, such as protein concentrations. These variables evolve over time, with the rate of change of each variable depending on the concentrations of others. Understanding these dynamics can help predict biological outcomes, such as the probability of apoptosis in a study of cancer, or provide insights into the temporal evolution of the system. Mathematically, such a system can be represented through the time-dependent variable concentrations $x(t)=(x_1(t), x_2(t), \ldots)$. The differential equations determining the time-evolution may be known through well-understood chemical or biological principles, or may need to be inferred through supervised learning based on experimental data.

We thus identify two initial goals for our framework:\\
a) Predict a set of discrete outcomes $y=(y_1, y_2, \ldots)$ given some set of concentration values $x=(x_1, x_2, \ldots)$\\
b) Given initial concentrations $x(t_0)=(x_1(t_0), x_2(t_0), \ldots)$, predict the concentrations at some future time $t$.

\subsection{Existing computational approaches}
There are several canonical machine learning approaches that can work on systems with a small to medium size number of features for predicting discrete biological outcomes. For systems at this scale, biological data scientists have had success with Python's scikit-learn module with built-in options for working with decision trees, support vector machines, random forest, and neural networks. However, as the number of features available increases with high throughput molecular assays, it can be difficult to obtain enough samples to match the number of measured features. When the number of samples is less than the number of features, the risk of over-fitting during the training of models increases, leading to a less generalizable solution. The other option is reduce the number of features utilized. This means possibly missing out on key information that drives subtle or non-linear effects on the emergent behavior of interest.

Neural networks have enjoyed considerable success over the last decade in making predictions using biological data with larger number of features. The main complaint about neural networks is their lack of transparency or 'explainabilty' in regards to identifying the mechanisms of action and key interactions driving the predicted result.

Mechanistic models spanning in size from individual signaling pathways to whole cell networks \cite{Erdem2022} provide explainability and the means to interrogate effects of precise peturbations of indivdual components. Many of these mechanistic models are based on ordinary differential equations representing biochemical reactions and protein-protein interactions. However as the size of these biological models grows in the number and types of features, so does the dimensionality and sparsity, making these models attractive targets to enhance through a hybrid quantum-classical approach.

\subsection{Quantum Algorithm Description}
\begin{figure}
    \centering
    \includegraphics[width=0.45\linewidth]{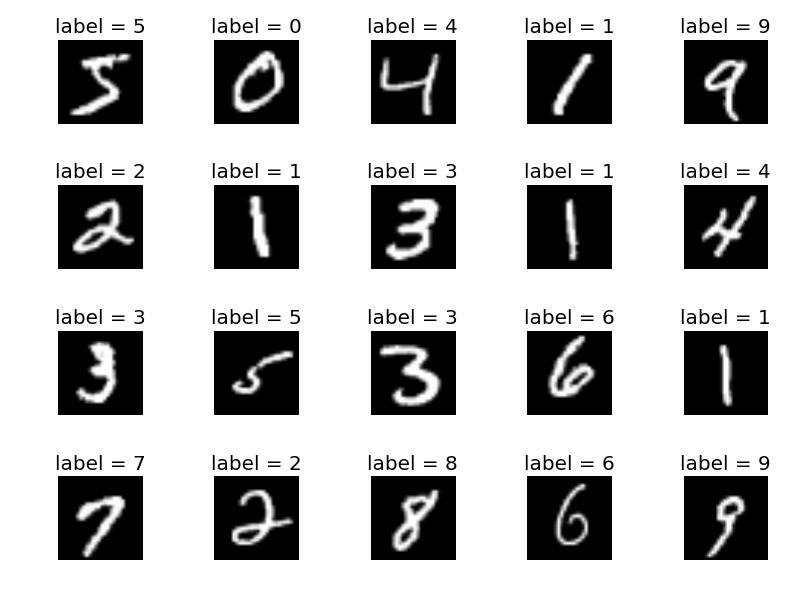}
    \hfill
    \includegraphics[width=0.5\linewidth]{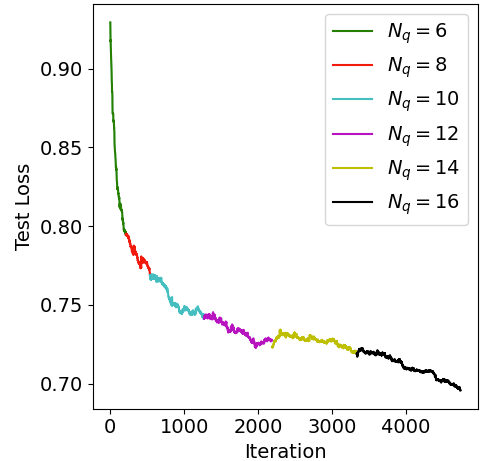}
    \caption{Illustration of training quantum models for classification with the encode-search-build approach on the MNIST dataset of handwritten digits (left). The curve (right) shows how the training loss decreases as the model training proceeds and how the loss decreases as models with larger number of qubits $N_q$ become available. Each color represents the performance of the model for a fixed number of qubits, with the training at each size beginning where the previous size concludes. The largest model with 16 qubits has 1407 parameters.}
    \label{fig:mnist}
\end{figure}

Our approach builds upon the techniques for training large QLMs presented in \cite{johri2025}. That work presented a scalable strategy for using quantum computers for classification. It presented a quantum classifier that encodes both the input and the output as binary strings which results in a model that has no restrictions on expressivity over the encoded data but requires fast classical compression of typical high-dimensional datasets to only the most predictive degrees of freedom. Second, it showed that if one parameter is updated at a time, quantum models can be trained without using a classical optimizer in a way that guarantees convergence to a local minimum, something not possible for classical deep learning models. Third, it proposed a parameter initialization strategy called sub-net initialization to avoid barren plateaus where smaller models, trained on more compactly encoded data with fewer qubits, are used to initialize models that utilize more qubits. Fig. \ref{fig:mnist} shows the training of QLMs using these techniques on up to 16 qubits for the 10 classes of the MNIST dataset.

\textit{QLM Design}: We first consider a supervised learning classification problem. We reproduce in brief here the discussion from \cite{johri2025} about loading classical data into a quantum state. We consider a dataset $\mathcal{D}=\{x,y\}$ which consists of data points $x=(x_1, x_2, \ldots)\in\mathcal{X}$ with accompanying labels $y\in\mathcal{Y}$, with the number of possible outputs $|\mathcal{Y}|\leq N_c$. Without loss of generality, we assume that $\mathcal{X}\in[0,1]^d$,  and $\mathcal{Y}_i\in\{0, 1, \ldots, N_c-1\}$. For a classification problem, the learning task consists of finding an efficiently computable function that maps $\mathcal{X}$ to $\mathcal{Y}$. 

Now let's find a binary encoding that approximates the original dataset as $\mathcal{D}_b=\{z_b,y\}$, where $b$ is a vector of natural numbers of dimension $d$ with $\sum_i b_i=B$. $z_{i}$ is the value of $x_i$ truncated to $b_i$ bits of precision, that is, $z_{ib} = \left\lfloor x_i 2^{b_i} \right\rfloor$. We include the case when $b_i=0$ for some components $i$. Henceforth, the argument $b$ will be dropped from $z$. As the magnitude of the $b_i$ grow, $\mathcal{D}_{b}$ more precisely captures the original dataset.

In this compressed dataset, there may be collisions at the input, so that a particular value of $z$ may occur more than once, each time corresponding to the same or different value of $y$. Then, the available data can be considered as samples from a joint probability distribution $f(\mathcal{Z}=z, \mathcal{Y}=y)$ over correlated random variables $(\mathcal{Z}, \mathcal{Y})$. For the purpose of classification, the `correct' output is then taken to be the mapping which occurs more frequently, that is 
\begin{align}
    C(z) = \arg \max_y f(z,y),
\end{align}
is the classification function to be learned.

\begin{figure*}
    \centering
    \includegraphics[width=\textwidth]{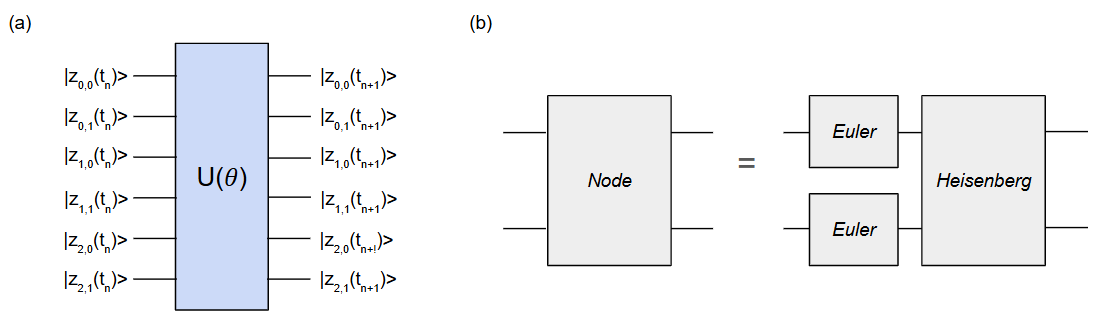}
    \caption{The architecture of the quantum model. (a) The circuit structure for modeling time evolution of variables in Eq. \ref{eq:bit_bit_unitary}. (b) The unitary $U$ consists of nodes that are entangling operations between qubits. Each node consists of single qubit operations parameterized by Euler rotations on each qubit and a Heisenberg-type interaction between the qubits.}
    \label{fig:quantum_model}
\end{figure*}

We see that $z$ can be exactly loaded into a quantum state that lives in the Hilbert space of $N_x \geq B$ qubits using Pauli $X$ gates acting on the $\ket{0}$ state to create the computational basis state $\ket{z}$. As $N_x$ grows, the representation of the dataset in the Hilbert space of the qubits becomes more exact. Similarly, the labels can be exactly mapped onto computational basis states $\ket{y}$ of $N_y=\left\lceil \log_2{N_c} \right\rceil$ qubits. Thus the classification function is of the form $C: \{0, 1\}^{N_x} \to \{0, 1\}^{N_y}$. We have thus approximated the problem of learning a multivariate function of real numbers to that of learning a vectorial Boolean function. Since both the input and output are now represented as binary strings, we call this a bit-bit encoding.

The quantum learning problem can now be cast in terms of a unitary $U_*$ that has the following action on $N_q=N_x+N_y$ qubits:
\begin{equation}
    \ket{0}\ket{z}\xrightarrow{U_*} \ket{C(z)}\ket{g(z)}
\end{equation}
where the states $\ket{g(z)}$ and $\ket{z}$ have support on $N_x$ qubits, and $N_y$ qubits initialized in state $\ket{0}$ are mapped by the unitary to the state $C(z)$. For the purpose of classification, we can discard the $g$ states. However, note that if $C(z)=C(z')$, we have $\langle g(z) | g(z')\rangle =\delta_{z,z'}$.

Thus, we can formulate the learning task on the quantum computer as that of finding a unitary $U$, parameterized by variables $\theta$, such that when it acts on $\ket{0}\ket{z}$, the probability of measuring $\ket{C(z)}$ at the output is maximized. That is, given,
\begin{align}
    U(\vec{\theta})\ket{0}\ket{z}=\sum_k \sqrt{P_{k,z}(\vec{\theta})}e^{i\phi_{k,z}(\vec{\theta})}\ket{k}\ket{g_{k,z}(\vec{\theta})},
\end{align}
find the value of $\vec{\theta}$ which minimizes the loss function
\begin{align}\label{eq:loss}
    \bar{L}(\vec{\theta})=1-\sum_{z\in \mathcal{D}_{\vec{b}}} f(z) P_{C(z),z}(\vec{\theta}),
\end{align}
where $f(z)$ is the frequency of occurence of $z$ in the dataset. $0\leq \bar{L}\leq 1$, and has the simple interpretation of being the probability of the model giving the wrong answer when queried. If $U(\vec{\theta})=U_*$ and there is no noise, even one shot would suffice to classify the input data sample.

Note that the loss function only includes the action of $U$ on values of $z$ in the dataset. The action of $U$ on values of $z$ not in the dataset will determine its generalization behavior. In that case, a quantum model that is perfectly trained on the dataset will output a probability distribution over the possible classes. 

We can extend the above construction with a quantum model that can also capture the evolution of correlated variables. At the input and output are concentrations encoded as bit strings.  In this case, the model unitary has the following action
\begin{equation}\label{eq:bit_bit_unitary}
    \ket{z(t_n)}\xrightarrow{U_*} \ket{z(t_{n+1})}
\end{equation}

The architecture of this QLM is shown in Fig. \ref{fig:quantum_model}. The unitary shown in Fig. \ref{fig:quantum_model} consists of several layers of entangling nodes and various designs of these layers will be explored with our platform. As the number of layers increases, the expressivity of the QLM will increase. Fig. \ref{fig:quantum_model} (b) shows the construction of each entangling node which acts on two qubits. It consists of single qubit Euler rotations acting on each qubit which correspond to unitaries in SU(2), followed by a `Heisenberg' unitary acting on both qubits. These are parameterized as follows:
\begin{align}\label{eq:node_def}
    U_{\text{Euler}}(\theta_1, \theta_2, \theta_3) = \text{Rx}(\theta_1) \text{Rz}(\theta_2) \text{Rx}(\theta_3)\nonumber\\
    U_{\text{Heisenberg}}(\theta_1, \theta_2, \theta_3) = \text{Rxx}(\theta_1) \text{Ryy}(\theta_2) \text{Rzz}(\theta_3)
\end{align}
where $\text{Rx}(\theta)$ is $\exp(-i\theta X)$ and $\text{Rxx}(\theta)$ is $\exp(-i\theta X\otimes X)$, and similarly for the other Pauli rotations. This construction is inspired by the general two-qubit gate decomposition in \cite{Vatan_2q}, and corresponds to a parameterized rotation of each qubit on its Bloch sphere before a parameterized entangling operation. An Euler unitary is also added to each qubit before it is measured.

The data encoding and circuit architecture discussed above results in a class of models that obey a form of \textit{universal approximation}, implying that there is no restriction on the functional form of the output with respect to the encoded data as the number of layers increases. In particular, this encoding can model arbitrary non-linear relationships between the encoded variables. This is in contrast to other data encoding schemes used in QLM such as angle or amplitude encoding, which enforce an inductive bias on the functional form of the model \cite{Schuld_data_encoding}.

\begin{figure}
    \centering
    \includegraphics[width=\linewidth]{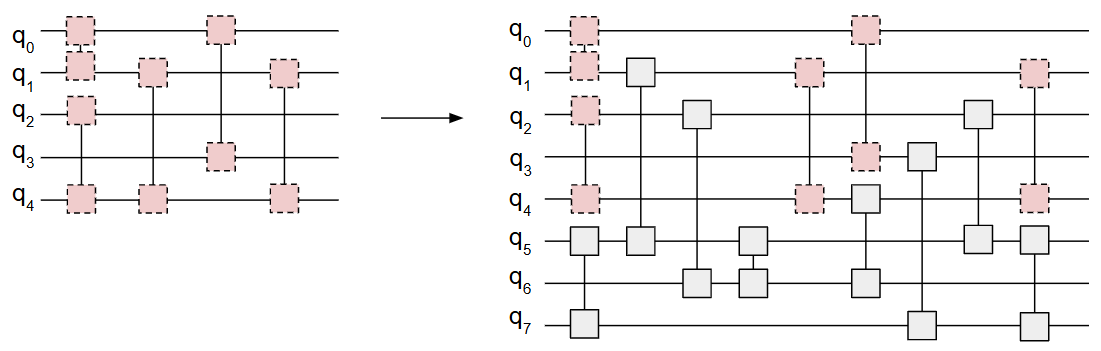}
    \caption{Example of an initialization from a sub-net taken from \cite{johri2025}. The model on the left has fewer parameters and acts on smaller input data. It is trained first and then used to initialize the corresponding sub-network in the model on the right (pink boxes, broken borders). Of the remaining parameterized nodes (gray boxes, solid borders), those connected to the sub-network are initialized as identity. The nodes not connected to the sub-network can be arbitrarily initialized. }
    \label{fig:subnet}
\end{figure}

Fig. \ref{fig:subnet} shows the sub-net initialization techniques presented in \cite{johri2025} which shows how the QLM can grow to include more variables as the quantum computational resources increase. For example, if a 40 qubit quantum computer is available this year that can support a requisite number of entangling gates, the parameters of a model trained on it can be used to initialize a larger model the next year when a 50 qubit quantum computer is available, and the training of the larger model does not need to start from scratch. In this way, training on near-term quantum computers can be viewed as a step towards training models on larger computers, and is thus of concrete value. Even if a model that can be trained today does not reach the baseline performance required to beat a classical model, it remains a crucial step in training a larger model anticipated to outperform classical counterparts. Thus, we can encode an expanding and hierarchical network of relationships between biological variables as quantum resources increase. To our knowledge, this is the first time such a clear argument has been presented for the direct utility of near-term quantum computers which are beyond the practical simulation capabilities of classical computers.

\textit{QLM Training}: For training QLMs, we will use the exact coordinate update scheme described in \cite{johri2025}. This scheme leverages a parameter shift rule to sequentially update the parameters of the model to iteratively refine model parameters with exact updates. Unlike classical deep learning models which have to be trained using approximate techniques like gradient descent, this scheme for training QLMs guarantees convergence to a local minimum and can avoid saddle points which plague many high-dimensional optimization problems.

\textit{Quantum Advantage}: With this encoding, quantum advantage in the time to correctly classify a new sample from a particular dataset will exist when $U_*$ can be approximated efficiently on a quantum computer but not classically. That is, with the quantum model, we aim to learn a function that is $\mathcal{O}(\exp(N_q))$ to compute classically while being $\mathcal{O}(\text{poly}(N_q))$ on a quantum computer. Eq. \ref{eq:bit_bit_unitary} encompasses unitaries based on Boolean functions which are the basis of many quantum algorithms with proven exponential speed-up such as Shor's algorithm for factoring. Most datasets however will not fit neatly into a compactly defined function, and often the only way to determine whether there is quantum advantage will be through actually training quantum models. 

An additional source of quantum advantage comes from the ability to train models to reach a minimum using the exact coordinate update technique discussed above. A third source comes from the lower energy cost per parameter of the QLM as compared to classical neural networks. This last advantage can be significant at scale \cite{pasqal-energy} but has not been studied systematically in the literature, and we intend to quantify this with our platform as well.

\section{Simulating Biological Abstractions with Quantum Circuits}
\begin{figure}
    \centering
    \includegraphics[width=\linewidth]{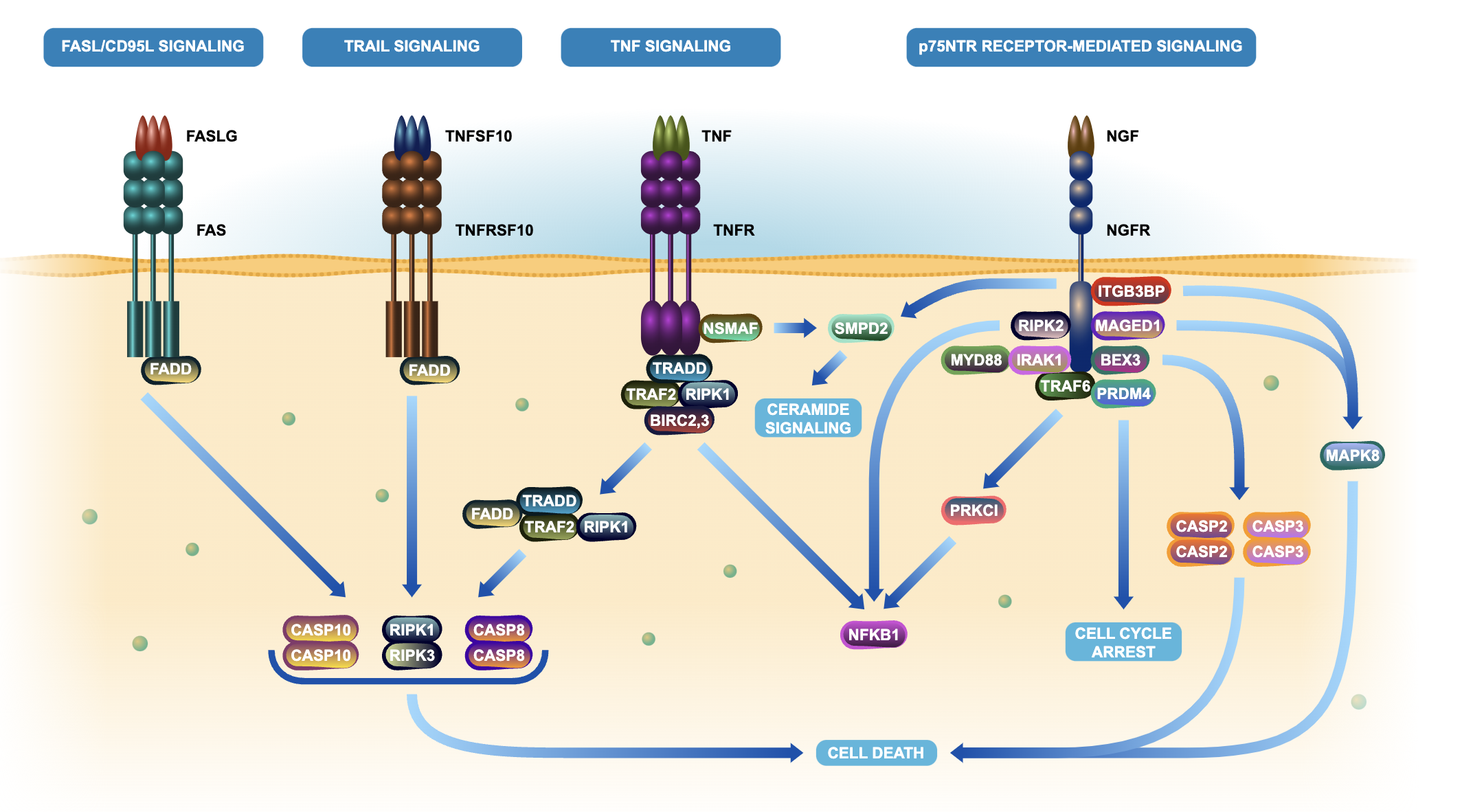}
    \caption{Key protein receptors and signaling interactions in apoptosis, programmed cell death, from \cite{Fabregat2018-hy}.}
    \label{fig:programmed cell death}
\end{figure}

As a proof of concept exercise, we explored the capability of our platform to predict if a cell was cancerous or healthy based on protein levels. Using data from 'The Human Protein Atlas' \cite{protein_atlas}, we extracted protein levels from tissues in 20 different types of cancer and 76 different cell types from healthy tissue. The extracted measurements came from proteins that mapped back to over 15k genes. Using protein level distributions across this data set we generated 100 cancer cells and 100 healthy cells. With the goal of building a quantum model that can form the basis for larger subnets in the future, we initially limited the scope of this initial model to 9 proteins in the Mangrum's and Finley's ODE model of apoptotic response of caspase-mediated in tumor cells \cite{MANGRUM2024111857} as visualized in Figure \ref{fig:programmed cell death} \cite{Fabregat2018-hy}.

The protein activities within a cell can be represented as a transform with inputs on one side being fed into a cascade of linked processes that result in the output of a distinct cellular state on the other side. Biological processes are mechanistic but also `squishy'. Their behaviors are often best modeled as probability distributions. These properties make our quantum circuit design a natural model for biological systems - the data are encoded into qubits as bit strings, the quantum circuit consisting of qubit rotations and entanglement implements the transform, and the aggregated, repeat measurements at the end gives a probability distribution of outcomes. The dependence of the output cellular state on the input protein levels encoded in the qubits at the start of the circuit can be adjusted by tuning the parameterized gates in the circuit. For example, a joint probability distribution over the binary levels of two proteins can be implemented using two qubits with a Controlled-RY gate, where the rotation parameter influences the correlation between the two proteins.



As the number of features in most biological datasets exceeds the encoding capacity of the number of qubits currently available on quantum simulators or hardware, a reduction in dimensionality is required. For example, in the experiment described here, each sample is associated with a vector of 9 real numbers that measure the levels of 9 different proteins. For a circuit of 4 qubits, each sample's vector will be compressed into a string of 3 bits for loading into the qubits, with one qubit reserved for reading out the binary classification outcome. With 3 qubits, only 8 different inputs are thus possible. However, the number of distinct biological states that can be represented increases exponentially with the number of qubits, quickly leading to comprehensive coverage of the input dataset. Here, the bit encoding is achieved through a reduction in dimensionality using principal component analysis (PCA) followed by a scheme that assigns different numbers of bits to each PCA direction based on its mutual information with the outcome \cite{johri2025}.


\begin{figure*}
    \centering
    \includegraphics[width=\textwidth]{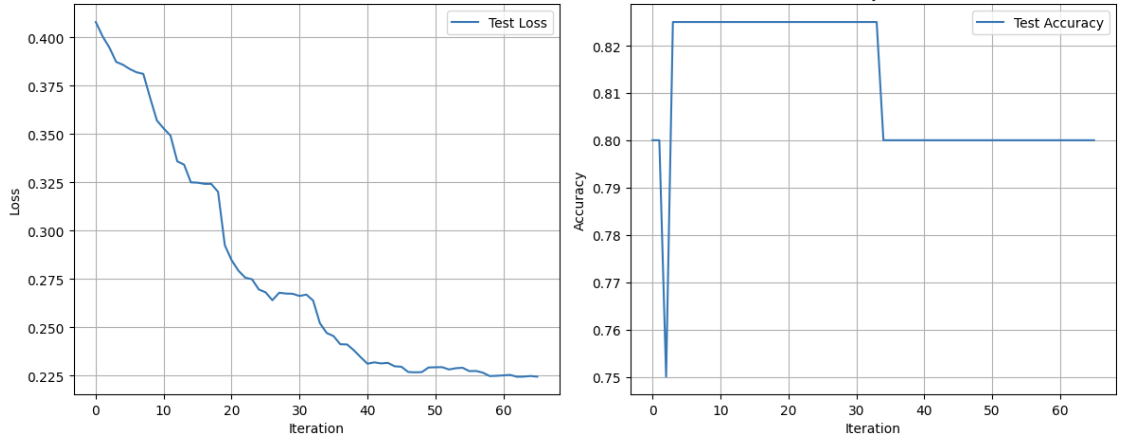}
    \caption{Test loss and accuracy performance with 4 qubits for the apoptosis classification problem described in Section IV using a dataset with 9 protein concentrations. While the loss consistently decreases, the accuracy is not monotonic, reflecting the imperfect correlation between the two.}
    \label{fig:9_protein_test}
\end{figure*}

For comparison to existing techniques, we utilized the same training and test datasets across the following models from scikit-learn with default settings: logistic regression, decision tree, random forest, support vector machine, multi-layer perceptron neural network. The achieved accuracies respectively were: 85\%, 75\%, 78\%, 88\%, 85\%. The maximum accuracy achieved by our platform with just 4 qubits was 82.5\% before the accuracy plateaued as shown in Fig. \ref{fig:9_protein_test}. This result serves as an initial validation of our approach. Increasing the number of proteins (model features) to 29, increases the accuracy of our quantum circuit to 95\% while still just using 4 qubits. This encouraging result shows that our compression and bit-encoding approach does not require the number of qubits to scale with the number of features, instead achieving an improved representation as the number of features increases even as the number of qubits stays constant. Further, increasing the number of qubits to 8 increases the accuracy on the 9 protein dataset to 87.5\%, validating the sub-net initialization approach.

By adopting a sub-net initialization approach, we can build accurate models for a small biological process, and then build more complex hierarchical models from these smaller models as the number of available qubits and measurements of multiple analytes (DNA, RNA, protein, metabolites, methylation, spatial omics) grow. The roadmap may go from a single-analyte pathway, to a multi-analyte pathway, to a multi-analyte network of pathways for a single cell, to an ecosystem of cells such as tumor microenvironment, to entire organ, to an entire organism. It also may evolve with DNA as foundational subnet, with regulatory and effector RNA, protein and epigenetic subnets getting stacked on top of one another.

\section{Feasibility and Impact}

Our platform will support building hybrid quantum-classical models for prediction of discrete states, simulating the evolution of states over time and modeling of interventions. Ultimately, we envision that decisions about interventions into human health can be informed by testing on a quantum-enabled digital twin. To realize this vision, existing bioinformaticians and data scientists must be able to effectively interface with advancements in quantum computing software and hardware. This will require a platform that serves as a user interface and middleware that agnostically bridges quantum resources with existing cloud-based machine learning,  bioinformatics and computational biology workflows. This will be achieved by connecting our platform with existing frameworks and data sets. 

For example, Navipoint Genomics \cite{navipoint} has a multicloud-based platform for running bioinformatics pipelines, and building and training machine learning and LLM models, as well as methods for securely connecting to NCI's Cancer Research Data Commons. We envision that NaviPoint Genomics users will be able to connect the rich data resources of the NCI's Cancer Research Data Commons to our platform using either our RESTful API or an optional Docker container deployment. With a RESTful API and dockerized deployment, we can enable integrations with several commercial (DNAnexus, Illumina's ICA, Velsera-Seven Bridges, TileDB) and academic groups (Terra and Galaxy) providing genomics informatics workflow and data warehousing solutions to life sciences organizations with valuable private-public data collections.      



This approach prioritizes accessibility, enabling bioinformaticians to leverage quantum advantages without requiring deep expertise in quantum mechanics. By providing an extensible architecture compatible with state-of-the-art quantum simulators and hardware, our platform addresses critical bottlenecks in scalability and usability. This work lays the foundation for quantum-driven discoveries in systems biology, offering a pathway to model complex emergent behaviors and accelerate therapeutic development.

While this paper focuses on quantum learning, the encode-search-build approach can also be extended to other computational problems in biology such as optimization or chemistry. We anticipate that our platform may also serve as a starting point for problems in these domains.

The primary roadblock to feasibility is potential delay in quantum hardware maturation timelines. As an example, IonQ, a manufacturer of trapped ion quantum computers, has projected that they will offer 64 Algorithmic Qubits by 2025, 256 by 2026 and 1024 by 2028 \cite{ionq_roadmap}. If the hardware timeline is delayed, the time to real-world quantum advantage using our platform will also correspondingly shift. However, the sub-net approach ensures that the models built during the extended timeline will still be of use, thus mitigating the risk factors of technological delays.


\section{Access and Collaboration}
Red Cedar is a quantum programming framework with a focus on machine learning and AI applications. It is currently under development at Coherent Computing Inc. Researchers interested in accessing the platform or collaborating on quantum omics applications are encouraged to contact the authors.

\newpage
\bibliographystyle{ieeetr}
\bibliography{references}

\end{document}